\documentclass[12pt]{iopart}

\usepackage{iopams}
\usepackage{graphicx}
\begin{document}

\title[CEWQO 17]{
Entanglement dynamics of bipartite system in squeezed vacuum
reservoirs}

\author{Smail Bougouffa$^{1}$ and Awatif Hindi$^{2}$}
\address{$^{1}$Department of Physics, Faculty of Science, Taibah University, P.O.Box 30002, Madinah, Saudi
Arabia\\ $^{2}$Physics Department, College of Science,  PO Box
22452, King Saud University, Riyadh - 11495, Saudi Arabia}
\ead{sbougouffa@taibahu.edu.sa\\ sbougouffa@hotmail.com}

\begin{abstract}
Entanglement plays a crucial role in quantum information
protocols, thus the dynamical behavior of entangled states is of a
great importance. In this paper we suggest a useful scheme that
permits a direct measure of entanglement in a two-qubit cavity
system. It is realized in the cavity-QED technology utilizing
atoms as flying qubits. To quantify entanglement we use the
concurrence. We derive the conditions, which assure that the state
remains entangled in spite of the interaction with the reservoir.
The phenomenon of sudden death entanglement (ESD) in a bipartite
system subjected to squeezed vacuum reservoir is examined. We show
that the sudden death time of the entangled states depends on the
initial preparation of the entangled state and the parameters of
the squeezed vacuum reservoir.

\end{abstract}

\pacs{03.67.Mn, 03.65.Yz, 03.65.Ud, 42.50.Lc}
\submitto{\PS}
\maketitle

\section{Introduction}\label{sec1}
In contrast to classical theories, quantum mechanics assures the
existence of nonlocal correlations between two systems spatially
separated and without any direct interactions, which Schrodinger
named as entanglement \cite{Sch35}. Entanglement is a key feature
of various quantum information processes such as quantum
teleportation \cite{BBCJPW93}, quantum dense coding \cite{BW92},
quantum cryptography \cite{E91} and quantum computing
\cite{BDEJ95}. Due to the crucial role of entanglement in quantum
information processes, the study of entanglement has attracted a
lot of interest in recent years. With various studies on
entanglement, the mean question which may be posed is how to know
that a quantum state is entangled. For a pure bipartite state, the
Schmidt decomposition \cite{S06} can be used to decide whether the
state is entangled and the degree of entanglement can be
quantified by the partial von Neumann entropy \cite{BBPS96}.
Hence, in principle, the problem of entanglement for pure states
of a bipartite system has been completely solved. On the other
hand, quantum systems predictably undergo decoherence processes
and quantum systems are mostly in mixed states. For density matrix
of a quantum system consisting of two subsystems, some criteria on
entanglement have been established
\cite{P96,HHH96,HH96,NK01,T00,S00,DGCZ00,GKLC01,HZ06,SV05}.
Moreover, the generation of entangled states has been investigated
in various systems from atoms or ions, photons and
quadrature-phase amplitudes of the electromagnetic field
\cite{HMNWBRH97,TWKMLIMW98,MS99,SKKLMMRTIWM00,DCZP00,JKP01,
RBONBRH01,FC72,C75,FT76,OM88,KMWZ95,BPDWZ98,PDGWZ01,ZCZYBP04,
THTS04,EKDSB04,BV04,ZWLJXP00}. It is clear that the experimental
and theoretical studies of bipartite systems have made a great
growth in recent
years.\\
Real quantum systems are necessarily subjected to their
environments, and these reciprocal interactions often result in
the dissipative evolution of quantum coherence and loss of useful
entanglement. Decoherence may be investigated in both local and
global dynamics, which may lead to the eventual deterioration of
entanglement \cite{ILZ07}. Yu and Eberly have investigated the
time evolution of entanglement of a bipartite qubit system
undergoing various modes of decoherence. Particularly, they found
that, even when there is no interaction, there are certain states
whose entanglement decays exponentially with time, while for other
closely related states, the entanglement vanishes abruptly in a
finite time which depends upon the initial preparation of the
qubits, a phenomenon termed entanglement sudden death
(ESD)\cite{YE07} and was recently observed in two sophistically
designed experiments with photonic qubits \cite{A07}and atomic
band \cite{L07}. Furthermore, it has also been observed in cavity
QED and trapped ion systems \cite{SMDZ06}. On the other hand, the
phenomenon ESD has motivated many theoretical investigations in
other bipartite systems involving pairs of atomic, photonic, and
spin qubits \cite{MLK08,G08,TPA05,CYH08}, multipartite systems
\cite{LRLSR08,S08} and spin chains \cite{CP08,LHMC08,AGXWL06}. In
addition, ESD has also been investigated for different
environments \cite{ ILZ07, YE07, AJ08,FT06,FT08}. However,
numerous investigations on ESD in a variety of systems have been
done so far, the question of ESD in interacting qubits remains
open \cite{DA09}. On the other hand, from the quantum
technological point of view, states that show exponential decay of
entanglement, and therefore maintain some trace of this all
considerable correlation for an infinite time, are of importance.
Although the vanishingly small entanglement present in the
exponential tail will be of limited practical importance, however
it is of interest to identify exactly in what situations ESD will
occur \cite{AJ08, S10}.\\
In the past few years, numerous methods by which the entanglement
of quantum systems can be detected and described have been
suggested. Possibly the mainly influence to date has been the
simple procedure derived by Wooters  \cite{W98} for measuring
entanglement for an arbitrary mixed state of pair two-level
systems. Furthermore, for two qubits, concurrence \cite{W98}
offers a convenient measure of the entanglement of formation. This
has provided a very useful tool for measurement of experimental
quantum states and is to day commonly used in evaluating the
abilities of emerging
quantum information technologies.\\
The purpose of this paper is to propose an efficient scheme for
quantum teleportation to generate entangled number states of
bipartite system under the influence of squeezed vacuum reservoir.
Thus we investigate the time evolution of these entangled states.
We examine the problem of ESD for this proposed scheme for
different initial entangled state and the parameters of the
squeezed vacuum reservoir.

\section{Bipartite model system}\label{sec2}
\begin{figure}
  \center{
  \includegraphics[height=0.4\textheight]{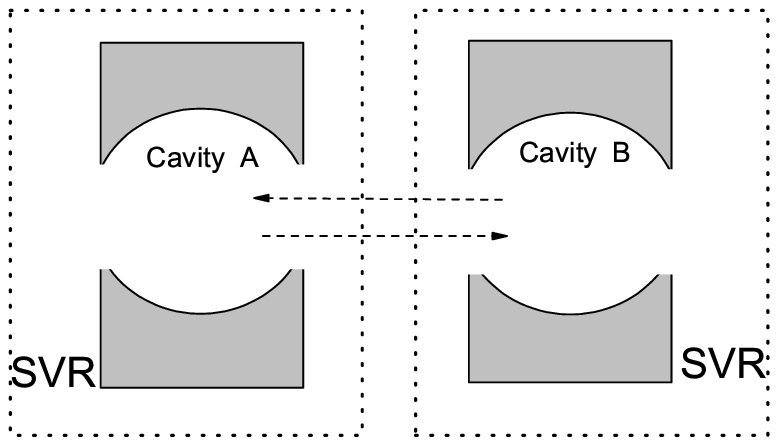}}
  \caption{Two independent systems of identical cavities containing
initial entangled fields. The entangled fields do not have
directional interaction with each other but independently interact
with their local environment in each cavity}\label{Fig0}
\end{figure}

Recently, Zubairy et all \cite{ZAS04} have suggested a new scheme
in their examination of the quantum disentanglement eraser. In
this simple scheme, the concurrence can be directly measured from
the visibility for an explicit class of entangled states. We
propose here the same scheme but with some adjustment. A two-level
atom with the upper level $|e\rangle$ and the lower level
$|g\rangle$ passes consecutively through cavity A, a squeezed
vacuum reservoir and a cavity B as shown in figure \ref{Fig0}. The
incident atom is initially prepared in the excited state
$|e\rangle$ and the decay of the radiation field inside a cavity
may be described by a model in which the mode of the field of
interest is coupled to a whole set of reservoir modes. We assume
that initially the two cavities are in vacuum state $|0\rangle$
and the atom always
leaves the setup in the ground state $|g\rangle$. \\
In the interaction picture and the rotating-wave approximation,
the Hamiltonian is simply
\begin{eqnarray}\label{1}
  H(t) = \hbar \sum_{j=A,B}\sum_{\mathbf{k}}\left[ g_{\mathbf{k}}^{(j)} b_{\mathbf{k}}^{(j)\dag}a_{j}
  e^{-i(\nu-\nu_{\mathbf{k}})t}+ g_{\mathbf{k}}^{(j)*}a_{j}^{\dag}b_{\mathbf{k}}^{(j)}
    e^{i(\nu-\nu_{\mathbf{k}})t}\right]
\end{eqnarray}
where $a_{j}(j=A,B)$ and $a^{\dag}_{j}$ are the destruction and
creation operators of the mode of the electromagnetic field of
frequency $\nu$. $b_{\mathbf{k}}^{j}$ and $b_{\mathbf{k}}^{\dag
j}$ are the modes of cavity j of frequency $\nu_{\mathbf{k}}$
which damp the field and $g_{\mathbf{k}}^{(j)}$ is the coupling
constant of the interaction between the electromagnetic field and
the cavity.

\section{Entanglement dynamics in squeezed reservoirs}\label{sec3}
Here we are concerned with the case in which cavity fields are
exposed in broadband squeezed vacuum reservoirs. From the general
analysis of system-reservoir interactions, when the modes
$b_{\mathbf{k}}^{j}$ are initially in a squeezed vacuum, with the
Hamiltonian (\ref{1}) and the squeezing bandwidths of the squeezed
reservoirs are much larger than the atomic line-widths, we can get
directly the master equation for the reduced density matrix for
the field in the cavities as \cite{SZ97}
\begin{eqnarray}\label{2}
  \dot{\rho}(t) = \sum_{j=A,B}\Bigg[&-&\frac{\kappa^{(j)}}{2}(N_{j}+1)\left(a_{j}^{\dag}a_{j}\rho(t)-
  2a_{j}\rho(t)a_{j}^{\dag}+\rho(t)a_{j}^{\dag}a_{j}
  \right)\nonumber\\
   &-&\frac{\kappa^{(j)}}{2}N_{j}\left(a_{j}a_{j}^{\dag}\rho(t)-2a_{j}^{\dag}\rho(t)a_{j}+
   \rho(t)a_{j}a_{j}^{\dag}
   \right)\nonumber\\
  &+& \frac{\kappa^{(j)}}{2}M_{j}\Big(a_{j}a_{j}\rho(t)-2a_{j}\rho(t)a_{j}+
   \rho(t)a_{j}a_{j}
   \Big)\nonumber\\
   &+&\frac{\kappa^{(j)}}{2}M_{j}^{*}\left(a_{j}^{\dag}a_{j}^{\dag}\rho(t)-2a_{j}^{\dag}\rho(t)a_{j}^{\dag}+
   \rho(t)a_{j}^{\dag}a_{j}^{\dag}\right)\Bigg]
\end{eqnarray}
where $\kappa^{(j)}(j=A,B)$ is the decay rate in the cavity,
$N_{j}=sinh^{2}(r_{j})$ and
$M_{j}=cosh(r_{j})sinh(r_{j})exp(-i\theta_{j})$, with $r_{j}$
being the squeeze parameter and $\theta_{j}$ being the reference
phase for the squeezed fields which surrounds the cavities A and
B. If $N_{j}=M_{j}=0$, the remaining terms are due to vacuum
fluctuations. \\
To investigate the effect of interaction among the bipartite on
decoherence we have to investigate the dynamics of bipartite
entanglement. Furthermore, the concept of concurrence initiates
from the original work of Hill and Wootters \cite{W98} where the
closed expression of the entanglement of formation of a system of
two qubits was derived. They established that the entanglement of
formation is a convex monotonic increasing function of the
concurrence. Here we use concurrence, to illustrate the degree of
entanglement for any bipartite system. This measure satisfies
necessary and sufficient condition for being good measure of
entanglement for 2X2 system. The concurrence varies from
$\mathcal{C}=0$ for a separable state to $\mathcal{C}=1$ for a
maximally entangled state. The explicit expression for concurrence
can be written as
\begin{equation}\label{3}
    C(t)=max(0,\sqrt{\lambda_{1}}-\sqrt{\lambda_{2}}-\sqrt{\lambda_{3}}-\sqrt{\lambda_{4}})
\end{equation}
where $\lambda'$s are the eigenvalues of the non-hermitian matrix
$\rho(t)\widetilde{\rho}(t)$ arranged in decreasing order of the
magnitude. The matrix $\rho(t)$ is the density matrix for the
bipartite and the matrix $\widetilde{\rho}(t)$ is given by
\begin{equation}\label{4}
    \widetilde{\rho}(t)=(\sigma_{y}^{A} \otimes\sigma_{y}^{B} )\rho^{*}(t) (\sigma_{y}^{A}\otimes
    \sigma_{y}^{B})
\end{equation}
where $\rho(t)^{*}$ is the complex conjugation of $\rho (t)$ and
$\sigma_{y}$ is the Pauli matrix given in quantum mechanics. In
the general case, we consider the field states in Fock basis in
two identical high-Q cavities A and B that represent a bipartite
system surrounding the entangled field as
\begin{equation}\label{5}
    |\Psi\rangle_{AB}(0)=\alpha_{1}|0_{A}0_{B}\rangle+\alpha_{2}|0_{A}1_{B}\rangle+
    \alpha_{3}|1_{A}0_{B}\rangle+\alpha_{4}|1_{A}1_{B}\rangle
\end{equation}
where $\alpha_{i} (i=1,2,3,4)$ are the probability amplitudes with
$\sum_{i=1}^{4}|\alpha_{i}|^{2}=1$. We use the basis (
$|1\rangle=|0_{A}0_{B}\rangle,|2\rangle=|0_{A}1_{B}\rangle,
|3\rangle=|1_{A}0_{B}\rangle,|4\rangle=|1_{A}1_{B}\rangle$) to
define the density matrix of the two qubit system. The equations
of motion in terms density matrix elements can be obtained using
the master equation \ref{2}.
\section{Results and conclusion}
Here we will consider some interesting initial entangled states
for the bipartite which can be prepared and have potential
applications in the quantum information processing tasks
\cite{S10}. We will begin by the examination of the EPR-states
which are perceptions in quantum information science, a vital part
of quantum teleportation and characterize the simplest possible
examples of entanglement.
\begin{enumerate}
\item Assume that the initially entangled state of the field in two cavities to be in a
NOON state given by
\begin{equation}\label{6}
|\Psi\rangle_{AB}(0)=\alpha|0_{A}1_{B}\rangle+
    \sqrt{1-\alpha^{2}}|1_{A}0_{B}\rangle
\end{equation}
This kind of state can be generated as it is mentioned in
\cite{S10} and having its potential application in
Heisenberg-limited metrology and quantum lithography \cite{B04}.
The solutions of the master equation for this initial
NOON state case are given in the Appendix A.\\

\item Consider now the initially entangled bipartite to be in a
another EPR-state given by
\begin{equation}\label{8}
    |\Psi\rangle_{AB}(0)=\alpha|0_{A}0_{B}\rangle+\sqrt{1-\alpha^{2}}|1_{A}1_{B}\rangle
\end{equation}
This kind of state can be prepared as we have mentioned in
\cite{S10}. States like these have been realized in experiments
with trapped ions \cite{L03}.
\end{enumerate}
The solution of the (Eq. 2) depends on the initial state of the
two bipartite system. We can show that, for these two classes of
the initial states that were be considered above, the solution of
(Eq.2) has the matrix shape in the representation spanned by the
two- bipartite states
\begin{equation}\label{81}
    \rho(t)=\left(%
\begin{array}{cccc}
  \rho_{11}(t) & 0 & 0 & \rho_{14}(t)\\
  0 & \rho_{22}(t) & \rho_{23}(t) & 0 \\
  0 & \rho_{32}(t) & \rho_{33}(t)& 0 \\
 \rho_{41}(t) & 0 & 0 & \rho_{44}(t) \\
\end{array}%
\right)
\end{equation}
With this form of the density matrix, we can show that the
concurrence can be expressed as
\begin{equation}\label{9}
    C(t)=max\left(0,\widetilde{C}_{1}(t),\widetilde{C}_{2}(t)\right)
\end{equation}
where
\begin{eqnarray}
  \widetilde{C}_{1}(t) &=& 2\left[\sqrt{\rho_{23}(t)\rho_{32}}-\sqrt{\rho_{11}(t)\rho_{44}(t)}\right] \\
   \widetilde{C}_{2}(t) &=&
   2\left[\sqrt{\rho_{14}(t)\rho_{41}}-\sqrt{\rho_{22}(t)\rho_{33}(t)}\right]
\end{eqnarray}
Using this formalism we can investigate the dynamics of
entanglement for the two initial states that considered above.
However, in the case of squeezed reservoirs, we find that the
entanglement sudden death always happens for the two initial
entangled states with $0<\alpha<1$. This is shown clearly in
numerical results plotted in Fig. 2. In Fig. 3, the time evolution
of the concurrence is plotted for different values of the degree
of squeezing. We note that, the sudden-death time of entanglement
becomes smaller as the degree of squeezed increases. In
conclusion, we investigate that, for bipartite entangled states,
the entanglement measured by concurrence abruptly disappears
during the dynamic evolution in the squeezed vacuum reservoir
while for the same class of entangled states, the entanglement
decays exponentially for vacuum reservoirs \cite{S10}. The results
can be extended to the high dimensional bipartite filed states
inside the cavities in squeezed vacuum environments(more than one
photon in each cavity)where the concurrence can not be used and we
have recourse to another measure of entanglement \cite{TIBZ10},
namely, the logarithmic negativity. The results are in Progress
and can be reported
elsewhere.\\

\verb Acknowledgments \\ The research of SB is supported by a
grant no (652/431) from the Deanship of Scientific Research,
Taibah University.

\newpage
\begin{figure}
  \center{
  \includegraphics[height=0.4\textheight]{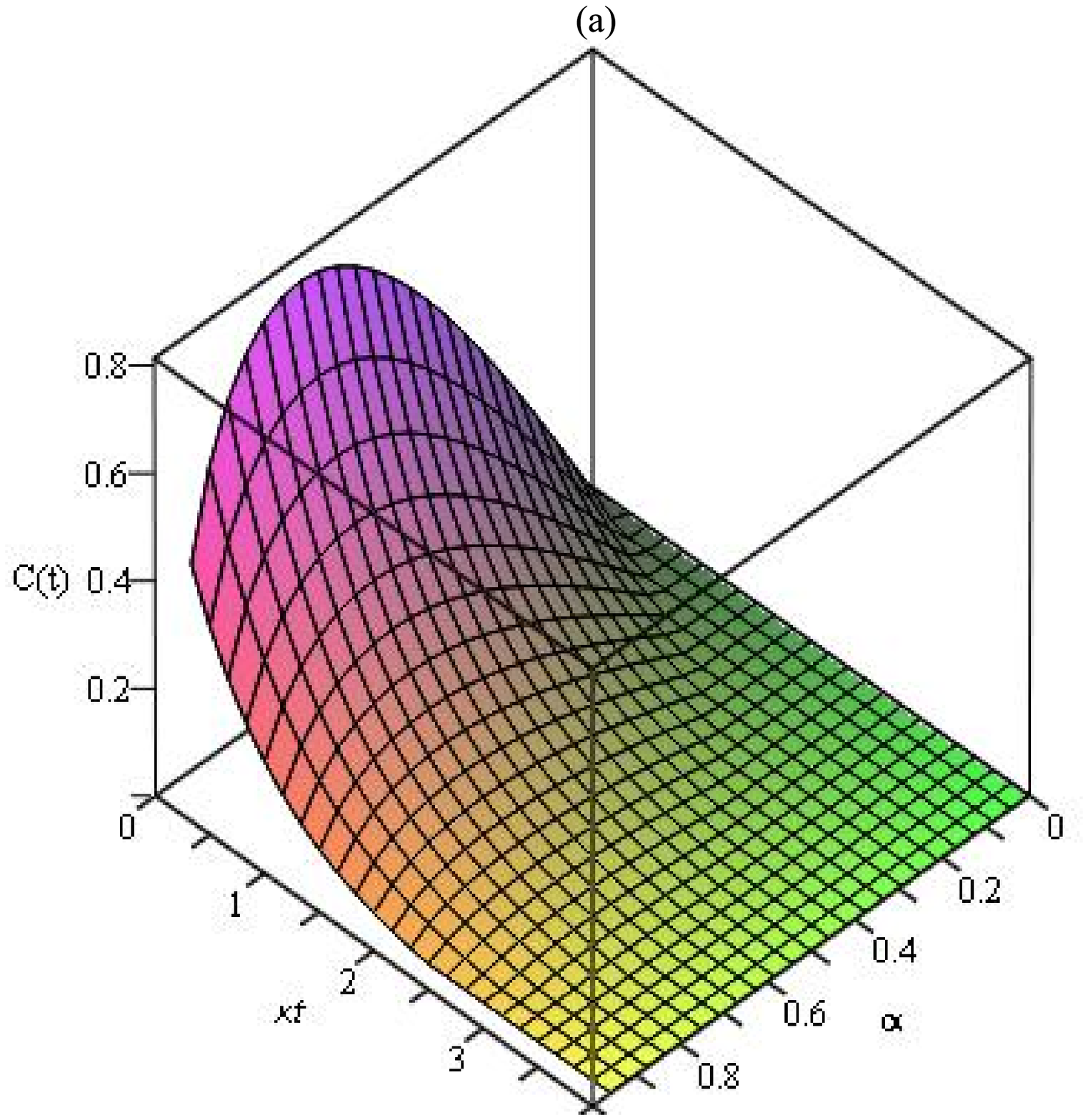}\\
  \includegraphics[height=0.4\textheight]{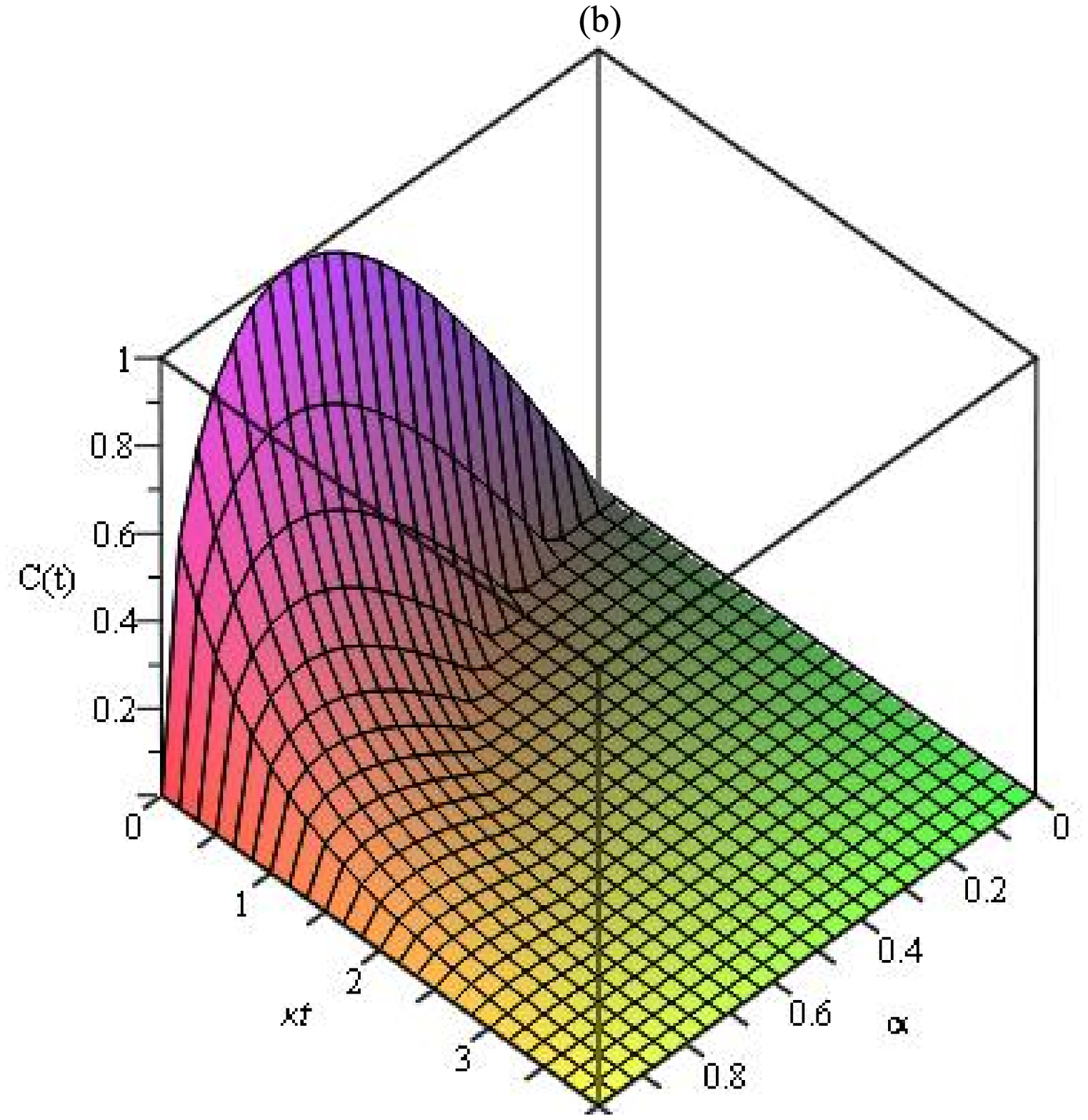}\\}
  \caption{(Color online)Entanglement dynamics of the two initial states of the bipartite system
  in squeezed vacuum reservoirs for $r=0.2$. (a) For the First initial NOON state. (b) For the second EPR initial state.
    }\label{Fig3}
\end{figure}
\begin{figure}
  \center{
  \includegraphics[height=0.4\textheight]{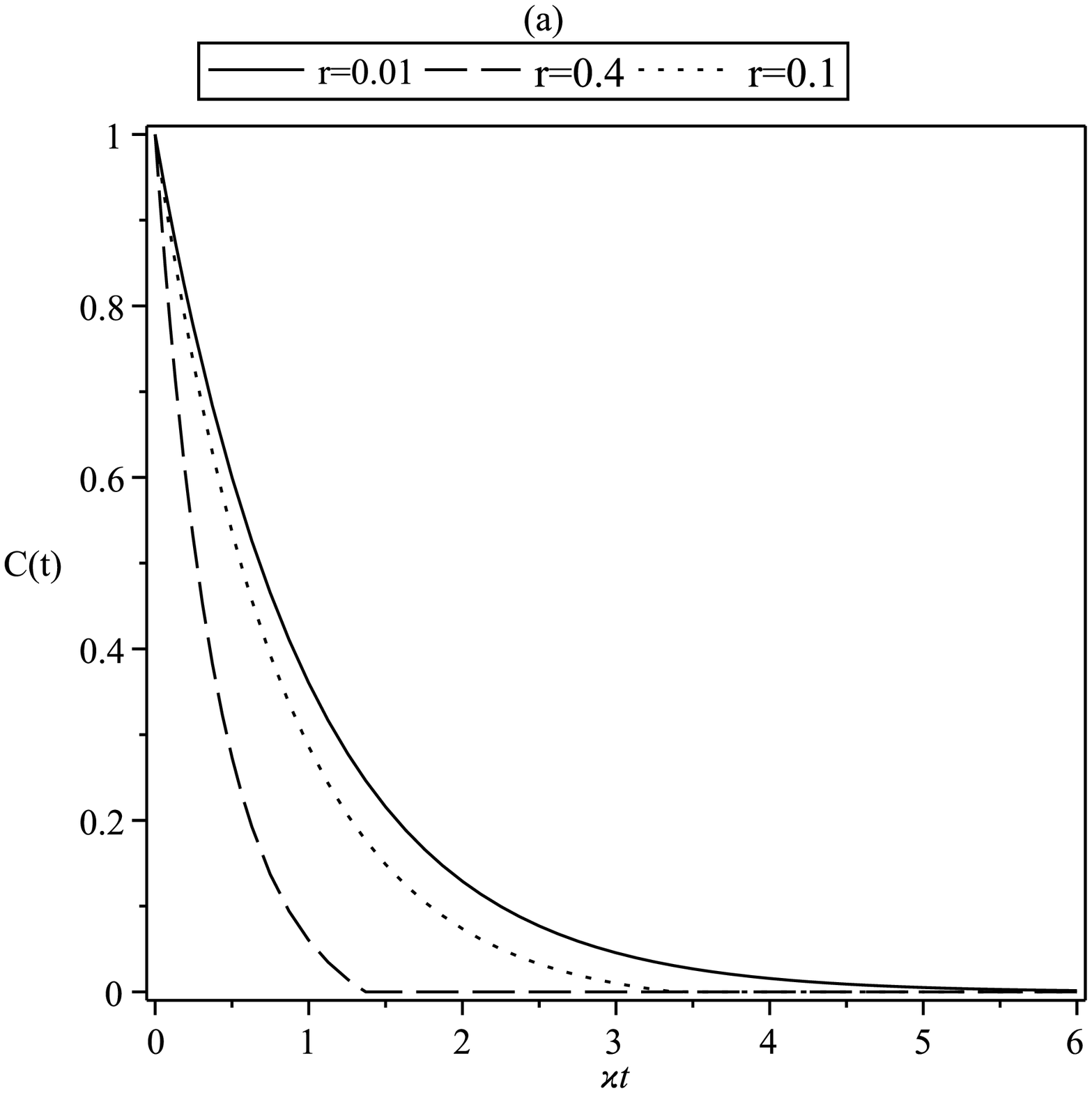}\\
  \includegraphics[height=0.4\textheight]{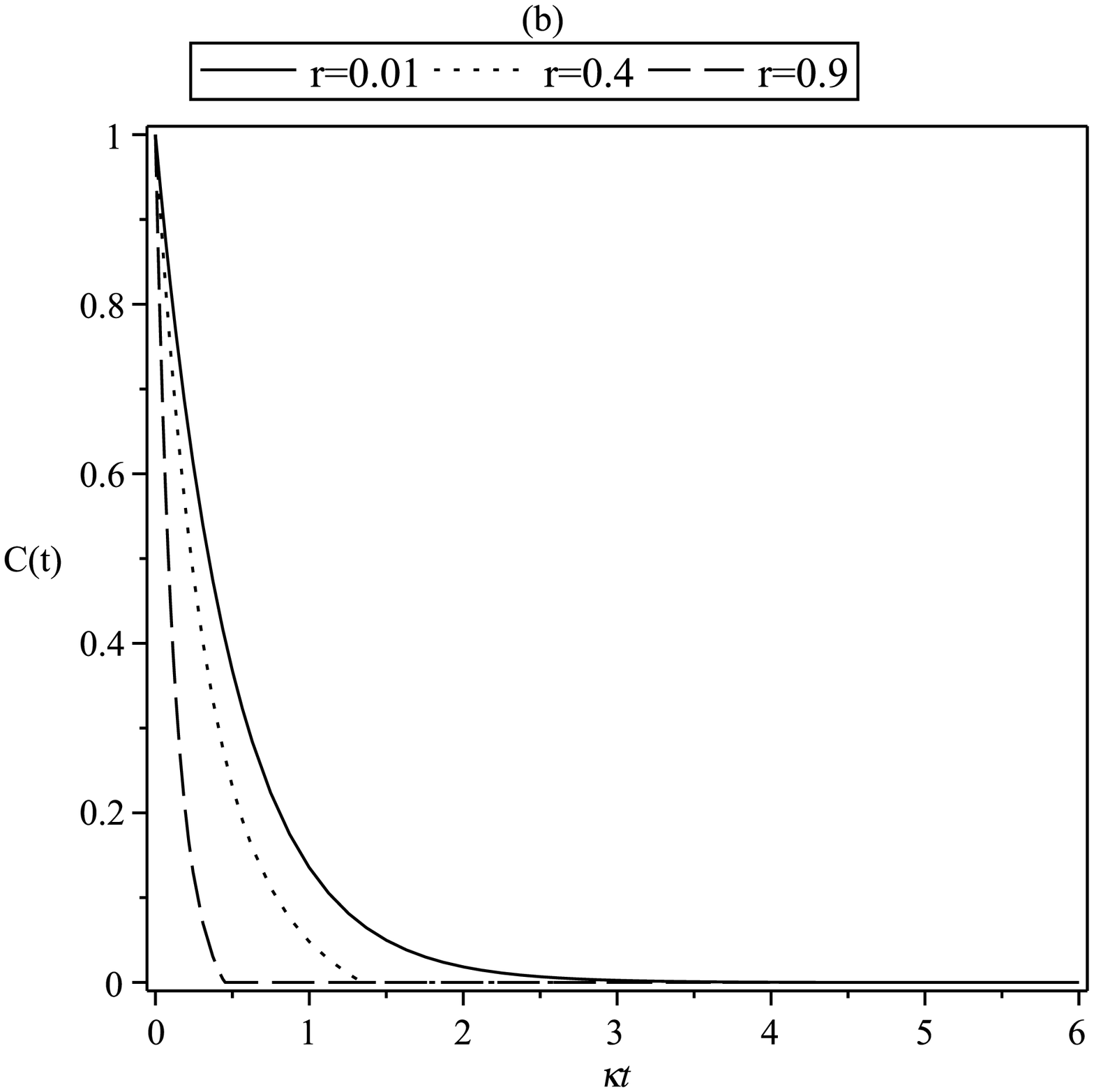}\\}
  \caption{Entanglement dynamics of the two initial states of the bipartite system
  in squeezed vacuum reservoirs for different values of the degree of squeezing and $\alpha=\frac{1}{\sqrt{2}}$.
  (a) For the First initial NOON state. (b) For the second EPR initial state.
    }\label{Fig3}
\end{figure}

\appendix
\section{Solutions of equations of motion of the density matrix elements for squeezed vacuum reservoirs}
The equation of motion of density matrix elements for the state
(Eq. 6) can be obtained and for the sake of simplicity, we assume
that the cavities are identical
$\kappa^{(A)}=\kappa^{(B)}=\kappa$, $N_{A}=N_{B}=N$ and
$M_{A}=M_{B}=M$ . On solving these equations of motion we get the
time evolution of the density elements matrix
\begin{eqnarray*}
\rho_{11}(t) &=& -\frac{a+3}{8b^2}\left[1+a+2\sinh(b\kappa t)-\frac{1+a}{b}\cosh(b\kappa t)\right]e^{-a\kappa t}\label{10}\\
\rho_{22}(t) &=& -\frac{1}{16}\left[(16\alpha^2+\frac{4}{b^2}-4)+(1-\frac{1}{b^2})\cosh(b\kappa t)\right]e^{-a\kappa t}\label{11} \nonumber\\
\rho_{33}(t)
&=&\frac{1}{16}\left[(-16\alpha^2+\frac{4}{b^2}+12)+(1-\frac{1}{b^2})\cosh(b\kappa
t)\right]e^{-a\kappa t}\label{12} \nonumber\\
\rho_{44}(t)
&=&\frac{1}{8b^2}\left[a^2-1+\frac{1}{4}(a-1)\Big(2b\sinh(b\kappa
t)-(a+1)\cosh(b\kappa t)\Big)\right]e^{-a\kappa t}\label{13}
\nonumber\\
\rho_{14}(t) &=&-\frac{M}{|M|}\alpha
\sqrt{1-\alpha^2}\sinh(|M|\kappa t)e^{-a\kappa
t}\nonumber\\\label{14} \rho_{32}(t) &=&\alpha
\sqrt{1-\alpha^2}\cosh(|M|\kappa t)e^{-a\kappa
t}\nonumber\label{15}
\end{eqnarray*}
and $\rho_{21}(t)=\rho_{12}^{*}(t)=0,
\rho_{31}(t)=\rho_{13}^{*}(t)=0, \rho_{32}(t)=\rho_{23}^{*}(t),
\rho_{41}(t)=\rho_{14}^{*}(t), \rho_{42}(t)=\rho_{24}^{*}(t)=0,
\rho_{43}(t)=\rho_{34}^{*}(t)=0$, where $a=4N+1$ and $
b^2=8N^2+8N+1$.

\end{document}